\documentclass[11pt,a4paper,twoside]{article}
\usepackage{cite}
\usepackage{graphicx}
\usepackage{epsfig}
\usepackage{dcolumn}% Align table columns on decimal point
\usepackage{bm}% bold math

\begin{document}

\def \d {{\rm d}}
\def \t {{\Theta}}
\def \k {{\kappa}}
\def \l {{\lambda}}
\def \s {{\sigma}}
\def \tr {{\tilde\rho}}
\def \tv {{\tilde v}}
\def \tz {{\tilde z}}
\def \e {{\epsilon}}
\def \P {{p_\lambda}}
\def \Q {{p_\nu}}

\newcommand{\be}{\begin{equation}}
\newcommand{\ee}{\end{equation}}

\newcommand{\beqn}{\begin{eqnarray}}
\newcommand{\eeqn}{\end{eqnarray}}
\newcommand{\AdS}{anti--de~Sitter }
\newcommand{\AAdS}{\mbox{(anti--)}de~Sitter }
\newcommand{\AAN}{\mbox{(anti--)}Nariai }
\newcommand{\AS}{Aichelburg-Sexl }
\newcommand{\pa}{\partial}
\newcommand{\pp}{{\it pp\,}-}
\newcommand{\ba}{\begin{array}}
\newcommand{\ea}{\end{array}}

\title{Robinson--Trautman spacetimes in higher dimensions}

\author{Ji\v{r}\'{\i} Podolsk\'y\thanks{Jiri.Podolsky@mff.cuni.cz} \ and Marcello Ortaggio\thanks{ortaggio`AT'science.unitn.it. Present address: Dipartimento di Fisica, Universit\`a degli Studi di Trento, and INFN, Gruppo Collegato di Trento, Via Sommarive 14, I-38050 Povo (Trento), Italy}
\\ Department of Theoretical Physics, Faculty of Mathematics and Physics,\\
 Charles University in Prague, V Hole\v{s}ovi\v{c}k\'{a}ch 2, 180 00 Prague 8,  Czech Republic}

\date{\today}

\maketitle

\abstract{As an extension of the Robinson--Trautman solutions of ${D=4}$ general relativity, we investigate higher dimensional spacetimes which admit a hypersurface orthogonal, non-shearing and expanding geodesic null congruence.  Einstein's field equations with an arbitrary cosmological constant and possibly an aligned pure radiation are fully integrated, so that the complete family is presented in closed explicit form. As a distinctive feature of higher dimensions, the transverse spatial part of the general line element must be a Riemannian Einstein space, but it is otherwise arbitrary. On the other hand, the remaining part of the metric is --- perhaps surprisingly --- not so rich as in the standard ${D=4}$ case, and the corresponding Weyl tensor is necessarily of algebraic type D. While the general family contains (generalized) static Schwarzschild--Kottler--Tangherlini black holes and extensions of the Vaidya metric, there is no analogue of important solutions such as the $C$-metric.}

\vspace{.2cm}

%%\keywords{Black Holes, Classical Theories of Gravity}

%\preprint{gr-qc/05XXXXX}

%\begin{document}

%\date{\today}

\section{Introduction}
\label{intro}

The Robinson--Trautman family of spacetimes, which was discovered almost fifty years ago \cite{RobTra60,RobTra62}, is one of the fundamental classes of exact solutions to Einstein's field equations (see chapter~28 in the comprehensive review book \cite{Stephanibook}). This large group of algebraically special spacetimes contains many important particular vacuum solutions, such as Schwarzschild-like black holes, accelerating black holes described by the $C$-metric, and other radiative spacetimes of various algebraic types. It also admits a cosmological constant, electromagnetic field or pure radiation, as in the special case of the Vaidya metric.

Geometrically, the Robinson--Trautman class is defined by admitting a geodesic, shear-free, twist-free but expanding null congruence. This invariant definition, based on optical scalars, can be applied to any theory of gravitation represented by a Lorentzian metric on a $D$-dimensional manifold. In the case $D=4$, the interplay between ``geometric optics'' and other properties of the gravitational field has been thoroughly explored (see \cite{Stephanibook} and original references therein). Among various remarkable results, the Goldberg-Sachs theorem states that a {\em vacuum} metric is algebraically special if and only if it contains a shear-free geodesic null congruence \cite{Stephanibook}. It is a result of resent research that the same proposition does not hold in higher dimension, and that shear comes to play an important role which is as yet not completely understood. For example, it has been demonstrated that the principal null directions of the $D=5$ Myers-Perry spacetime (which is of type D) are shearing \cite{FroSto03,Pravdaetal04}. Furthermore, in vacuum spacetimes of type III or N, a multiple principal null direction with expansion necessarily has also non-zero shear \cite{Pravdaetal04}. This implies in particular that there do not exist Robinson-Trautman vacuum solutions of type III or N when $D>4$, as opposed to the $D=4$ case \cite{Stephanibook}. In this context, the knowledge of explicit Robinson-Trautman metrics in higher dimension would be desirable. It is the purpose of the present paper to systematically derive the $D>4$ Robinson--Trautman class of solutions and to discuss its main features. Here we confine ourselves to the most important case of vacuum spacetimes, with a possible cosmological constant, but we allow for the presence of aligned pure radiation. In section~\ref{sec_geom} we study the general form of the line element under purely geometrical requirements on the optical scalars. In section~\ref{Einstein} we derive the explicit solutions to Einstein's equations within such a setting. We summarize and discuss the obtained spacetimes in section~\ref{sec_discussion}, and present concluding remarks in section~\ref{sec_conclusions}. Appendix~A discusses the structure of the Weyl tensor of the general metric, and appendix~B provides some useful identities.

\section{Geometrical assumptions}

\label{sec_geom}

Throughout the paper we will basically follow the strategy of \cite{RobTra62}. Given a $D$-dimensional spacetime ($D\ge4$), let us consider a family of null hypersurfaces $u(x)=const$, i.e. with normal (and tangent) $k_\alpha=- u_{,\alpha}$ satisfying $g^{\alpha\beta}k_\alpha k_\beta=0$. This automatically guarantees that the null congruence of integral curves of the vector field $k^\alpha=g^{\alpha\beta}k_\beta$ is {\em geodesic} and {\em affinely parametrized}, i.e. $k_{\alpha;\beta}k^\beta=0$. It is natural to take the function $u$ itself (constant along each ray) as one of the coordinates, so that $k_\alpha=-\delta^u_\alpha$ and $g^{uu}=0$. As for the remaining coordinates, we use the affine parameter $r$ along the geodesics generated by $k^\alpha$, and ``transverse'' spatial coordinates $(x^1, x^2, \ldots , x^{D-2})$ which are constant along these null geodesics. This further implies $k^\alpha=\delta^\alpha_r$, that is $g^{ur}=-1$ and $g^{uu}=0=g^{ui}$.
Therefore, the covariant line element can be written as
\be
 \d s^2=g_{ij}\left(\d x^i+ g^{ri}\d u\right)\left(\d x^j+ g^{rj}\d u\right)-2\,\d u\d r-g^{rr}\d u^2 ,
 \label{geo_metric}
\ee
where the metric functions may depend arbitrarily on all the coordinates $(x,u,r)$; from now on, $x$ stands for all the transverse coordinates $x^i$ and lowercase latin indices ranges as $i=1,\ldots,D-2$.
Useful relations between the covariant and contravariant metric coefficients, to be employed in the sequel, are
\be
 g^{ri}= g^{ij}g_{uj} , \qquad g^{rr}=-g_{uu}+g^{ij}g_{ui}g_{uj} , \qquad g_{ui}= g^{rj}g_{ij} ,
 \label{cov_contra}
\ee
while $g_{rr}=0=g_{ri}$.
In this coordinate system, it is also easy to see that
\be
 k_{\alpha;\beta}=\frac{1}{2}g_{\alpha\beta,r} .
\ee
For later purposes, it is convenient to define an auxiliary $(D-2)$-dimensional spatial metric $\gamma_{ij}$ by
\be
 \gamma_{ij}=p^2g_{ij} , \qquad p^{2(2-D)}\equiv\det g_{ij}=-\det g_{\alpha\beta} ,
 \label{gamma and P}
\ee
so that $\det\gamma_{ij}=1$. Then one can express the generalized optical scalars \cite{FroSto03,Pravdaetal04} associated to $k^\alpha$ simply as\footnote{Using the standard identity $\gamma^{ij}\gamma_{ij,r}=\left[\ln (\det\gamma_{ij})\right]_{,r}$ which implies $\gamma^{ij}\gamma_{ij,r}=0$. Notice that the twist matrix \cite{Pravdaetal04} identically vanishes here, i.e. $k_{(\alpha;\beta)}=k_{\alpha;\beta}$, due to the assumed hypersurface orthogonality of $k^\alpha$. It is also worth emphasising that the definitions of the scalars $\theta$ and $\sigma^2$ in~(\ref{scalars}) hold only when an {\em affine} parameter is used along~$k^\alpha$.}
\beqn
 & & \theta\equiv\frac{1}{D-2}k^\alpha_{\;;\alpha}=-(\ln p)_{,r} , \nonumber \label{scalars} \\
 & & \sigma^2\equiv k_{(\alpha;\beta)}k^{\alpha;\beta}-\frac{1}{D-2}(k^\alpha_{\;;\alpha})^2=\frac{1}{4}\gamma^{li}\gamma^{kj}\gamma_{ki,r}\gamma_{lj,r} ,
\eeqn
where $\theta$ and $\sigma$ are, respectively, the expansion and the shear scalars.

Now, imposing the additional condition that the congruence $k^\alpha$ is {\em shear-free}, $\sigma^2=0$, (which is equivalent to the vanishing of the full shear matrix $\sigma_{li}$ of \cite{Pravdaetal04}), eq.~(\ref{scalars}) leads to
\be
 \gamma_{ij,r}=0 ,
\ee
since there always exists a frame in which $\gamma^{ij}$  is diagonal, with strictly positive eingenvalues.

To summarize, the line element of any spacetime admitting a hypersurface orthogonal (therefore geodesic) shear-free null congruence $\mbox{\boldmath$k$}=\pa_r$ can thus be written in the form (\ref{geo_metric}),
with $g_{ij}=p^{-2}\gamma_{ij}$; the matrix $\gamma_{ij}$ is unimodular and independent of $r$, while $p$, $g^{ri}$ and $g^{rr}$ are arbitrary functions of $(x,u,r)$.

Such a metric is left invariant by the following coordinate transformations (which do not change the family of null hypersuperfaces $u=const$ nor the affine character of the parameter~$r$):
\be
 x^i=x^i(\tilde x,\tilde u) , \qquad u=u(\tilde u), \qquad r=r_0(\tilde x,\tilde u)+\tilde r/\dot u(\tilde u) .
 \label{freedom}
\ee
Since any two-dimensional metric is conformally flat, when $D=4$ one can always use such coordinate freedom to cast the metric $\gamma_{ij}$ into an explicitly conformally flat form \cite{RobTra62}. However, for $D>4$ this is generally not the case, so that our approach will now slightly depart from that of Robinson and Trautman. Nevertheless, we shall recover the results of \cite{RobTra62} in the special case $D=4$ (see the end of subsection~\ref{subsec_Ruu}).

\section{Einstein's field equations}

\label{Einstein}

The next step is to discuss specific consequences of plausible physical requirements, in particular imposing Einstein's equations with a suitable energy-momentum tensor in the above Robinson--Trautman class. In the present paper we concentrate on spacetimes which do not contain matter fields, except (possibly) {\em pure radiation} aligned with the null vector $k_\alpha$, characterized by $T_{\alpha\beta}=\Phi^2 k_\alpha k_\beta$. In the coordinate system introduced above this means that only the $T_{uu}$ component can be non-vanishing. In addition, we admit the presence of an arbitrary {\em cosmological constant} $\Lambda$. Our analysis will in particular reduce to investigation of {\em vacuum} spacetimes when both $T_{uu}=0$ and $\Lambda=0$.

Recalling the line element~(\ref{geo_metric}) and the relations~(\ref{cov_contra}), the full Einstein field equations $R_{\alpha\beta}-\frac{1}{2}Rg_{\alpha\beta}+\Lambda g_{\alpha\beta}=8\pi T_{\alpha\beta}$ thus reduce to the following set of equations: $R_{rr}=0$, $R_{ri}=0$, $R_{ij}=\frac{2}{D-2}\Lambda g_{ij}$, $R_{ur}=-\frac{2}{D-2}\Lambda$, $R_{ui}=\frac{2}{D-2}\Lambda g_{ui}$, and $R_{uu}=\frac{2}{D-2}\Lambda g_{uu}+8\pi \Phi^2$. We now proceed to integrate all such equations.

\subsection{The equation $R_{rr}=0$}

\label{subsec_Rrr}

Since we have assumed that twist and shear are identically vanishing, a generalized Sachs equation\footnote{Namely, $(D-2)\theta_{,\alpha} k^\alpha+\sigma^2+(D-2)\theta^2+\mbox{Tr}(A^2)=-R_{\alpha\beta}k^\alpha k^\beta$, where $A$ is the twist matrix \cite{Pravdaetal04}. This identity follows from the definition of the Riemann tensor applied to any geodesic null vector $k^\alpha$ with an affine parameter. Cf.~also \cite{LewPaw05}, and eq.~(6.33) of \cite{Stephanibook} for $D=4$.} implies that for the $R_{rr}$ component we have simply
\be
 R_{rr}=-(D-2)(\theta_{,r}+\theta^2) .
\ee The integration of the field equation $R_{rr}=0$ singles out
two alternative solutions. The first arises when $\theta=0$ which,
using eq.~(\ref{scalars}), is equivalent to $p=p(x,u)$. This
possibility corresponds to the Kundt class of spacetimes admitting
a non-expanding null congruence, some properties of which in $D>4$
have been discussed, e.g., in \cite{ColHerPel06} and references
therein. Alternatively, for $\theta\not=0$ the vector field
$k^\alpha$ has {\em non-zero expansion} and one finds
$\theta^{-1}=r+r_0(x,u)$. By analogy with the well-known $D=4$
case, we refer to this as the Robinson--Trautman class of
spacetimes: it will be the subject of this paper. The arbitrary
function $r_0$ can always be set to zero by a
transformation~(\ref{freedom}). From eq.~(\ref{scalars}) we can
thus factorize  $p=r^{-1}P(x,u)$, where $P$ is an arbitray
function. It is also convenient to rescale the transverse spatial
metric by introducing $h_{ij}=P^{-2}\gamma_{ij}$, so that the
relation  (\ref{gamma and P}) can now be written as
\be
 g_{ij}=r^2h_{ij}(x,u) .
 \label{hspatial}
\ee
This specifies the $r$-dependence of $g_{ij}$ in the metric (\ref{geo_metric}).

\subsection{The equation $R_{ri}=0$}

The next Einstein equation to consider is $R_{ri}=0$. After some calculations one finds
\be
 R_{ri}=-\frac{1}{2}h_{ij}\,r^{2-D}\left(r^D g^{rj}_{\;\;\; ,r}\right)_{,r} .
\ee
The field equation thus requires
\be
 g^{ri}=e^i(x,u)+r^{1-D}f^i(x,u) ,
  \label{gri}
\ee
where $e^i$ and $f^i$ are arbitrary functions of $x$ and $u$, while the $r$-dependence of the metric functions $g^{ri}$ is now completely fixed.

\subsection{The equation $R_{ij}=\frac{2}{D-2}\Lambda g_{ij}$}

Somewhat lengthier calculations give the following ``transverse'' Ricci tensor components
\beqn
 R_{ij}= & & {\mathcal R_{ij}}-h_{ij}\left[r^{4-D}\left(r^{D-3}g^{rr}\right)_{,r}+rg^{rk}_{\;\;\; ,k}+rg^{rk}(\ln\sqrt{h})_{,k}-r(\ln\sqrt{h})_{,u}\right]  \\
 & & {}-\frac{1}{2}r^{4-D}\left[r^{D-2}\left(2h_{k(i}{g^{rk}}_{,j)}
+g^{rk}h_{ij,k}\right)\right]_{,r}-\frac{1}{2}r^4g^{rk}_{\;\;\; ,r}g^{rl}_{\;\;\; ,r}h_{ik}h_{jl}
+\frac{D-2}{2}rh_{ij,u} , \nonumber
\eeqn
where
\be
h \equiv \det h_{ij} \qquad\hbox{so that}\qquad \sqrt{h}=P^{2-D}\,, \label{h}
\ee
${\mathcal R_{ij}}$ is the Ricci tensor associated with the spatial metric $h_{ij}$, and indices within small round brackets are symmetrized. We wish now to solve the equation $R_{ij}=\frac{2\Lambda}{D-2} g_{ij}$. Using previous eqs.~(\ref{hspatial}) and (\ref{gri}), one can split it in terms with different $r$-dependence. By inspection, it follows that $g^{rr}$ must be of the form
\be
 g^{rr}=c_1+c_2r+c_3r^2+c_4r^{2-D}+c_5r^{2(2-D)}+c_6r^{3-D} ,
 \label{grr_old}
\ee
where $c_1,\ldots, c_6$ are arbitrary functions of $(x,u)$. Substituting this expression into $R_{ij}=\frac{2\Lambda}{D-2} r^2h_{ij}$ (and contracting with $h^{ij}$ when necessary) determines five of such six functions as
\beqn
 & & c_1=\frac{{\cal R}}{(D-2)(D-3)} , \qquad c_2=\frac{2}{D-2}\left[(\ln\sqrt{h})_{,u}-e^k_{\;,k}-e^k (\ln\sqrt{h})_{,k}\right] ,   \nonumber \label{c_i} \\
 & & c_3=-\frac{2\Lambda}{(D-1)(D-2)} , \qquad c_4=\frac{D-3}{D-2}\left[f^k_{\;,k}+f^k (\ln\sqrt{h})_{,k}\right] , \\
 & & c_5=\frac{1}{2}\frac{D-1}{D-2}h_{kl}f^k f^l ,\nonumber
\eeqn
where ${\cal R}=h^{ij}{\mathcal R_{ij}}$.
In addition, one finds that the metric $h_{ij}$ and the functions $e^i$ and $f^i$ must satisfy the constraints
\beqn
 {\mathcal R}_{ij}&=&\frac{{\mathcal R}}{D-2}h_{ij} , \nonumber \\
2h_{k(i}{e^k }_{,j)} +e^k h_{ij,k}-h_{ij,u}&=&\frac{2}{D-2}\left[e^k_{\;,k}+e^k (\ln\sqrt{h})_{,k} -(\ln\sqrt{h})_{,u}\right]h_{ij} , \label{constrij} \\
 2h_{k(i}{f^k }_{,j)} +f^k h_{ij,k}&=&\frac{2}{D-2}\left[f^k_{\;,k}+f^k (\ln\sqrt{h})_{,k}\right]h_{ij} \nonumber ,\\
(h_{ik}f^k)(h_{jl}f^l)&=&\frac{1}{D-2}(h_{kl}f^k f^l)\,h_{ij} \nonumber .
\eeqn

First, it is easy to see that the left-hand side of the last equation for $f^i$ represents a matrix with zero determinant. Thus, the determinant on the right-hand side must also vanish. Since $h_{ij}$ is positive definite and thus non-degenerate, the only possible solution is that $h_{kl}f^k f^l=0$, which necessarily implies
\be
 f^i=0 .
 \label{f^i}
\ee This automatically satisfies the third equation
in~(\ref{constrij}). In addition, thanks to eq.~(\ref{f^i}) we can
use the coordinate freedom~(\ref{freedom}) to remove the
coefficients $g^{ri}=e^i$ in the metric~(\ref{geo_metric}). Thus,
without loss of generality we may set (at least locally)
\be
 e^i=0 .
 \label{e^i}
\ee
Explicitly, this can be achieved by performing the transformation
\be
x^i=x^i(\tilde x, u) \equiv -\int e^i(\tilde x,u)\,\d u.
\ee
This is regular only when the determinant $\det J^i_{\,j}$ of the Jacobi matrix
${J^i_{\,j}=\frac{\partial x^i}{\partial\tilde x^j}}$ is non-vanishing. In the degenerate case $\det J^i_{\,j}=0$, we may alternatively remove $e^i$, for example, with the transformation
\be
x^i= -\int e^i(\tilde x,u)\,\d u - \lambda\,\tilde x^i \label{tran2} .
\ee
This is now clearly regular provided $\lambda\not=0$ is any real parameter different from the eigenvalues of the matrix $J^i_{\,j}$ in the neighbourhood of a given point.

To summarize, dropping tildes, we are now left with the Robinson--Trautman metric in the form
\be
 \d s^2=r^2h_{ij}\,\d x^i\d x^j-2\,\d u\d r-g^{rr}\d u^2 ,
 \label{geo_metric spec}
\ee
where, using eqs.~(\ref{grr_old}), (\ref{c_i}), (\ref{f^i}), (\ref{e^i}), the coefficient $g^{rr}$ is explicitly given by,
\be
g^{rr}=\frac{{\cal R}}{(D-2)(D-3)}+\frac{2(\ln\sqrt{h})_{,u}}{D-2}\,r-\frac{2\Lambda}{(D-2)(D-1)}\,r^2-\frac{\mu}{r^{D-3}} \,.
\label{grr}
\ee
The function $\mu(x,u)$, which renames~$c_6$, is arbitrary, and the $(D-2)$-dimensional spatial metric $h_{ij}$
is constrained by the first two equations in~(\ref{constrij}) with $e^i=0$, namely
\beqn
 {\mathcal R}_{ij}&=&\frac{{\mathcal R}}{D-2}h_{ij} , \label{constrijr} \\
  h_{ij,u}&=&\frac{2}{D-2} h_{ij}(\ln\sqrt{h})_{,u} . \label{constrijs}
\eeqn
As a well-known fact, any two-dimensional metric $h_{ij}$ satisfies the first equation (\ref{constrijr}), so that this is identically satisfied in the special case $D=4$. As a particular consequence, for $D=4$ the scalar curvature ${\mathcal R}$ of $h_{ij}$ will generally depend both on $u$ and on the spatial coordinates~$x$. On the other hand, for any $D>4$, eq.~(\ref{constrijr}) tells us\footnote{Just compare the covariant derivative of eq.~(\ref{constrijr}) with the contracted Bianchi identities (in the geometry of $h_{ij}$) to get ${\mathcal R}_{,j}=0$ (unless $D=4$). We shall return to this specific point shortly from a different viewpoint.} that ${\mathcal R}$ can depend only on the coordinate $u$, and that at any given $u=u_0=const$ each spatial metric $h_{ij}(x,u_0)$ must describe an {\em Einstein space} (for $D=5$ this implies that $h_{ij}(x,u_0)$ corresponds to a 3-space of constant curvature).

The latter equation (\ref{constrijs}) ``controls'' the parametric dependence of $h_{ij}(x,u)$ on $u$, and it can be easily integrated to obtain $h_{ij}=h^{1/(D-2)}\,\gamma_{ij}(x)$. Consequently, $ h\equiv \det h_{ij}=h \det \gamma_{ij}$, so that the matrix $\gamma_{ij}$ must be unimodular. Considering eq.~(\ref{h}) we can write
\be
h_{ij}=\frac{\gamma_{ij}(x)}{P^2(x,u)} \qquad\hbox{where}\qquad \det \gamma_{ij}=1\, \label{hijs}
\ee
(note that $\gamma_{ij}$ coincides with the $\gamma_{ij}$ matrix introduced in section~\ref{Einstein}).
The spatial metric $h_{ij}(x,u)$ can thus depend on the coordinate $u$ only via the conformal factor $P^{-2}$.

\subsection{The equation $R_{ur}=-\frac{2}{D-2}\Lambda$}
The next Ricci tensor component for the metric (\ref{geo_metric spec}) reads
\beqn
 R_{ur}= & & \frac{1}{2}r^{2-D}\left(r^{D-2}g^{rr}_{\;\;\; ,r}\right)_{,r} -r^{-1}(\ln\sqrt{h})_{,u}\,  .
\eeqn
By substituting the expression (\ref{grr})  we obtain $R_{ur}=-\frac{2}{D-2}\Lambda$, so that the corresponding field equation is automatically satisfied.

\subsection{The equation $R_{ui}=\frac{2}{D-2}\Lambda g_{ui}$}
The Ricci tensor component $R_{ui}$ for the metric (\ref{geo_metric spec}) is
\beqn
 R_{ui}&=& \frac{1}{2}r^{4-D}\left(r^{D-4}g^{rr}_{\;\;\; ,i}\right)_{,r}
  + \frac{1}{2}\left(h^{jk}h_{ik,u}\right)_{,j} \nonumber\\
 && {}+\frac{1}{2}h^{jk}h_{ik,u}(\ln\sqrt{h})_{,j} -\frac{1}{4}h^{jk}h^{lm}h_{kl,u}h_{jm,i}-(\ln\sqrt{h})_{,ui}\, .
\eeqn
Now we use relation (\ref{constrijs}), substitute for $g^{rr}$ from eq.~(\ref{grr}) and we employ the field equation $R_{ui}=\frac{2}{D-2}\Lambda g_{ui}=0$ (cf. eq.~(\ref{geo_metric spec})). Comparing the coefficients of different powers of $r$ we obtain two simple conditions
\be
(D-4)\,{\cal R}_{,i}=0 ,\qquad\quad  \mu_{,i}=0 . \label{Ri}
\ee
It immediately follows that the function $\mu$ must be independent of the spatial coordinates,
\be
 \mu=\mu(u),
\ee
and similarly
\be
 {\cal R}={\cal R}(u)\ \hbox{ for } D>4 . \label{RD}
\ee
In the unique case $D=4$ the first equation in (\ref{Ri}) is satisfied identically, so that one can have a much more general function ${\cal R}(x,u)$.  This explicitly confirms our arguments presented above which were based on the application of the Bianchi identities in relation to constraints  (\ref{constrijr}).

\subsection{The equation $R_{uu}=\frac{2}{D-2}\Lambda g_{uu}+8\pi \Phi^2$}

\label{subsec_Ruu}

Finally, we evaluate the Ricci tensor component $R_{uu}$. It turns out that
\beqn
 R_{uu}&= & \frac{1}{2}g^{rr}g^{rr}_{\;\;\; ,rr}-\frac{1}{2}g^{rr}_{\;\;\; ,r}(\ln\sqrt{h})_{,u}
 +\frac{D-2}{2r}\left(g^{rr}g^{rr}_{\;\;\; ,r}+g^{rr}_{\;\;\; ,u}\right) \nonumber\\
&& {}+\frac{1}{2r^2}\left[\left(g^{rr}_{\;\;\;
,i}h^{ij}\right)_{,j}+g^{rr}_{\;\;\;
,i}h^{ij}(\ln\sqrt{h})_{,j}\right]
-(\ln\sqrt{h})_{,uu}-\frac{1}{4}h^{il}h^{jk}\,h_{ij,u}h_{kl,u}\,
.\label{Ruu} \eeqn When we substitute the explicit
form~(\ref{grr}) of the metric function $g^{rr}=-g_{uu}$  (cf.
eq.~(\ref{cov_contra})), we use the constraint (\ref{constrijs})
and the relation $\sqrt h=P^{2-D}$, see (\ref{h}), we obtain \beqn
 R_{uu} &=& \frac{2}{D-2}\Lambda g_{uu}+\frac{D-2}{2\,r^{D-2}}\,\left[ (D-1)\mu(\ln P)_{,u}-\mu_{,u} \right]\nonumber\\
&& {}+\frac{1}{2(D-2)(D-3)\,r^2}\,\left[ ({\mathcal
R}_{,i}h^{ij})_{,j}-(D-2)({\mathcal R}_{,i}h^{ij})(\ln P)_{,j}
\right] \label{Ruun}\\ && {}+\frac{1}{r}\left[ \frac{{\mathcal
R}_{,u}-2{\mathcal R}(\ln P)_{,u}}{2(D-3)} -\big(h^{ij}(\ln
P)_{,ui}\big)_{,j}+(D-2)h^{ij}(\ln P)_{,ui}(\ln
P)_{,j}\right].\nonumber \eeqn

This must now satisfy $R_{uu}=\frac{2}{D-2}\Lambda g_{uu}+8\pi \Phi^2$. The first term in eq.~(\ref{Ruun}) is thus exactly of the required form, and can be cancelled. In addition, the last term (proportional to $r^{-1}$) vanishes  identically. Indeed, since $h_{ij}=P^{-2}\gamma_{ij}$ (cf.~eq.~(\ref{hijs})) and $\left(\gamma^{ij}(\ln P)_{,i}\right)_{,j}={}^{\gamma}\Delta(\ln P)$ (where $^{\gamma}\Delta$ is the Laplace operator in the geometry of $\gamma_{ij}$), such term can be rewritten as
\be
\frac{1}{2r}\left[\frac{{\mathcal R}_{,u}-2{\mathcal R}(\ln P)_{,u}}{D-3}-P^2\left[2\,{}^{\gamma}\Delta(\ln P)-(D-4)\gamma^{kl}(\ln P)_{,k}(\ln P)_{,l}\right]_{,u}\right] ,
\ee
which equals zero thanks to eq.~(\ref{diff_scalar}) in appendix~B. We can now concentrate on the remaining two terms proportional to $r^{2-D}$ and $r^{-2}$.

The Bianchi identities imply $T^{\alpha\beta}_{\quad ;\beta}=0$ which, using the first definition in eq.~(\ref{scalars}), for $T_{\alpha\beta}=\Phi^2 k_\alpha k_\beta$ gives the equation $(\ln\Phi^2)_{,r}=(2-D)\,\theta$. In particular, for the Robinson--Trautman family of spacetimes $\theta=r^{-1}$ (cf. subsection~\ref{subsec_Rrr}), so that the possible pure radiation is characterized by the function
\be
\Phi^2=r^{2-D}n^2(x,u)\,,
\ee
where $n$ is an arbitrary function of $x$ and $u$. This allows us to easily compare the terms with different powers of $r$ in the field equation for $R_{uu}$. In higher dimensional spacetimes with $D>4$ one has ${\mathcal R}_{,i}=0$, see (\ref{RD}), so that we have a non-vanishing contribution only from the $r^{2-D}$ term. We thus obtain the condition
\be
 (D-1)\,\mu\,(\ln P)_{,u}-\mu_{,u} =\frac{16\pi\, n^2}{D-2} \qquad (D>4) \,. \label{constrspec2}
\ee

On the other hand, in the exceptional case $D=4$ the two terms proportional to $r^{2-D}$ and to $r^{-2}$ are both non-vanishing and, obviously, with the same $r$ dependence. Hence they sum up to give the more complicated equation
\be
 \frac{1}{2}({\mathcal R}_{,i}h^{ij})_{,j}-({\mathcal R}_{,i}h^{ij})(\ln P)_{,j}+6\mu(\ln P)_{,u}-2\mu_{,u} =16\pi n^2 \qquad (D=4) \,.
 \label{RTD4}
\ee
The first two terms represent (one-half of) the covariant Laplace operator on a 2-space with metric $h_{ij}$ (applied to ${\mathcal R}$). Renaming $\mu=2m$ we thus obtain
$ \frac{1}{2}\Delta{\mathcal R}+12m(\ln P)_{,u}-4m_{,u}=16\pi n^2$, which is the familiar form of the Robinson--Trautman equation (see eq.~(28.71) in \cite{Stephanibook}). Let us emphasise that in the $D=4$ canonical form given in \cite{RobTra62,Stephanibook}, the 2-dimensional spatial metric takes the conformally flat form $h_{ij}=P^{-2}(x,u)\delta_{ij}$. In fact, for $D=4$ we can achieve this by a transformation $x^i=x^i(\tilde x)$ involving only the spatial coordinates $x$, since the $u$-dependence of the spatial metric is factorized out as in eq.~(\ref{hijs}).

\section{Discussion of Robinson--Trautman solutions in $D>4$}

\label{sec_discussion}

We can thus present the complete explicit form of the Robinson--Trautman spacetimes (aligned pure radiation or vacuum with a cosmological constant $\Lambda$) in any dimension $D>4$ as
\be
 \d s^2=\frac{r^2}{P^2}\,\gamma_{ij}\,\d x^i\d x^j-2\,\d u\d r-2H\,\d u^2 ,
 \label{geo_metric fin}
\ee
where the function $2H\equiv g^{rr}=-g_{uu}$ is  given by
\be
2H=\frac{{\cal R}(u)}{(D-2)(D-3)}-2\,r(\ln P)_{,u}-\frac{2\Lambda}{(D-2)(D-1)}\,r^2-\frac{\mu(u)}{r^{D-3}} \,.
\label{Hfin}
\ee
The unimodular spatial metric $\gamma_{ij}(x)$ and the function $P(x,u)$ must satisfy the field equations~(\ref{constrijr}) for the Einstein metric $h_{ij}=P^{-2}\gamma_{ij}$. Having thus established the function $P(x,u)$, we can prescribe an arbitrary function $\mu(u)$ and then employ  the last eq.~(\ref{constrspec2})
--- which is the equivalent of the Robinson--Trautman equation in higher dimensions ---  to evaluate the corresponding function $n^2(x,u)$ which characterizes pure radiation in the spacetime.

Notice that using the coordinate freedom remaining from (\ref{freedom}), namely the reparametrization of $u$,
\be
u=u(\tilde u),\   r=\tilde r/\dot u(\tilde u), \quad \hbox{so that} \quad \tilde P=P\,\dot u\,,\ \tilde{\cal R}={\cal R}\,{\dot u}^2\,,\ \tilde \mu=\mu\,{\dot u}^{D-1}\,,
 \label{freedomspec}
\ee
($\tilde{\cal R}$ is indeed the Ricci scalar of the rescaled metric $\tilde h_{ij}={\dot u}^{-2}h_{ij}$) we may achieve further useful simplification of the metric (\ref{geo_metric fin}), (\ref{Hfin}). For example, we can always put $\tilde\mu$ to be a constant. Alternatively, in $D>4$ we can set $\tilde{\cal R}$ to a constant.
Also, transformations of the coordinates $x^i=x^i(\tilde x)$ can be used to change the form of the spatial metric $h_{ij}$.  In any case, for fixed $r$ and $u$ the metric $\tilde h_{ij}$ can be any positive definite Einstein space, see (\ref{constrijr}), the $u$-dependence of this family being governed solely by $P$.

We have thus found that for the family of the Robinson--Trautmane spacetimes in higher dimensions the metric function $H$, given explicitly by (\ref{Hfin}), is simple. In fact, it is much simpler than in the case $D=4$ because ${\cal R}$ can not depend on $x$. In this sense, the Robinson--Trautman class admits a \emph{richer structure in four dimensional general relativity} than in higher dimensions (see also the sequel for further comments). On the other hand, the variety of possible spatial metrics $h_{ij}$ is huge. For example, if ${\cal R}>0$ and $5\le D-2\le 9$ one can take any of the infinite number of non-trivial compact Einstein spaces presented and classified in \cite{Bohm98}. One can also easily verify, see Appendix~A, that {\em higher dimensional Robinson--Trautman vacuum/pure radiation solutions (with an arbitrary cosmological constant) can be only of type D} (or conformally flat, i.e. type O, in which case they reduce to constant curvature spacetimes) in the algebraic classification of \cite{Coleyetal04}. Again, this should be contrasted with the case $D=4$, in which all algebraically special types are allowed.

For certain purposes it may be useful to rewrite the crucial equation (\ref{constrijr}) explicitly for $\gamma_{ij}$ and $P$. Using the conformal relation $h_{ij}=P^{-2}\gamma_{ij}$ and eq.~(\ref{tracefree}), eq.~(\ref{constrijr}) becomes
\beqn
 \frac{1}{D-4}\left({}^{\gamma}{\mathcal R}_{ij}-\frac{{}^{\gamma}{\mathcal R}}{D-2}\gamma_{ij}\right)= & &
 -(\ln P)_{||ij}-(\ln P)_{,i}(\ln P)_{,j} \nonumber \label{hg} \\
 & & {}+\frac{1}{D-2}\,\gamma_{ij}\left[{}^{\gamma}\Delta(\ln P)+\gamma^{kl}(\ln P)_{,k}(\ln P)_{,l}\right] ,
\eeqn
where the lower double vertical bar denotes the covariant derivative with respect to $\gamma_{ij}$.

The simplest class of Robinson--Trautman spacetimes in higher dimensions obviously arises when the Einstein metric $h_{ij}$ is of {\em constant curvature} and thus conformally flat (note that this is automatically true when $D=5$, because $h_{ij}$ is then three-dimensional). This implies that with a suitable transformation of the coordinates $x^i$ we can always set
\be
\gamma_{ij}=\delta_{ij}. \label{flat}
\ee
In this case, eq.~(\ref{hg}) simplify enormously and can be explicitly integrated to obtain the function $P(x,u)$. Namely, we have ${P_{,ij}=0}$ for ${i\not=j}$, which imply that the coordinates $x^i$ are separated, i.e. $P$ must be of the form  $P=P_1(x^1,u)+P_2(x^2,u)+\cdots$. The relation $P_{,11}=P_{,22}=\cdots$ then tells us that there is a common separation function $c(u)$. The complete function $P(x,u)$ can thus be integrated as
\be
P=a(u)+b_i(u)\,x^i+c(u)\,\delta_{ij}x^ix^j, \label{flatP}
\ee
where $a(u), b_i(u), c(u)$ are arbitrary functions of $u$. The corresponding Ricci scalar in (\ref{Hfin}) is
\be
\frac{{\cal R}}{(D-2)(D-3)}=4ac-\sum_{i=1}^{D-2} b_i^2.
\label{sect}
\ee
This gives the general explicit form of the Robinson--Trautman pure radiation spacetimes~(\ref{geo_metric fin}) with a spatial metric $h_{ij}=P^{-2}\delta_{ij}$ of constant curvature. These also contain the $D>4$ counterpart of the well-known Vaidya metric \cite{Stephanibook}.

\subsection{Vacuum solutions}

\label{subsec_vacuum}

When the pure radiation field vanishes, i.e. $n=0$, we obtain \emph{vacuum} Robinson--Trautman spacetimes (\ref{geo_metric fin}) in higher dimensions with a general Einstein metric $h_{ij}=P^{-2}\gamma_{ij}$. In this case, one has to consider separately the two cases $\mu\neq 0$ and $\mu=0$.

When $\mu\neq 0$, it can always be set to a constant by a rescaling~(\ref{freedomspec}). Eq.~(\ref{constrspec2}) thus reduces to ${\,\mu P_{,u}=0}$. Consequently, in this case the function $P$ must be independent of the coordinate $u$, and thus must therefore be also $h_{ij}$. It follows, in particular, that ${\cal R}$ is a constant. Unless now ${\cal R}=0$, one can choose the transformation~(\ref{freedomspec}) to set ${\cal R}=\pm (D-2)(D-3)$. The corresponding vacuum Robinson--Trautman solutions are thus fully characterized by the line element~(\ref{geo_metric fin}) with the simple function
\be
 2H=K-\frac{2\Lambda}{(D-2)(D-1)}\,r^2-\frac{\mu}{r^{D-3}} \, , \qquad (K=0, \pm 1) \, ,
\label{Hvacuum}
\ee
and they clearly admit $\pa_u$ as a Killing vector. As mentioned above, the spatial metric $h_{ij}=P^{-2}(x)\gamma_{ij}(x)$ can describe {\em any} Einstein space with scalar curvature ${\cal R}=K(D-2)(D-3)$. When this space is compact, such a family describes various well-known black hole solutions. In particular, if the horizon has constant curvature one obtains Schwarzschild--Kottler--Tangherlini black holes  \cite{Tangherlini63}, for which the line element can always be cast in the form\footnote{In this case, with the flat spatial metric (\ref{flat}) the function $P$ is given by  (\ref{flatP}) in which $a,b_i,c$ are now constants. With a linear transformation of $x$ we can put $P$ into a canonical form as in eq.~(\ref{geo_metric spec2}).}
\be
 \d s^2=r^2\left(1+\textstyle{\frac{1}{4}}K\,\delta_{kl}x^kx^l\right)^{-2}\,\delta_{ij}\,\d x^i\d x^j-2\,\d u\d r-2H\,\d u^2 .
 \label{geo_metric spec2}
\ee
In addition, there are generalized black holes \cite{Birmingham99,GibIdaShi02prl,GibHar02} with various horizon geometries (e.g., those presented in \cite{Bohm98} for $K=+1$), non-standard asymptotics and, possibly, non-spherical horizon topology. All these solutions are of type D (see appendix~A). In passing, this also proves that the static black ring of \cite{EmpRea02prd} does not belong to the Robinson--Trautman class, since it is of type I$_i$ \cite{PraPra05}.

In the exceptional case in which $\mu=0$, eq.~(\ref{constrspec2}) is identically satisfied, and one can not conclude that $P$ is independent of $u$. One can still rescale ${\cal R}(u)$ to be a constant ${\cal R}=K(D-2)(D-3)$,
$K=0, \pm 1$, so that in this case the line element~(\ref{geo_metric fin}) contains the characteristic function
\be
 2H=K-2\,r(\ln P)_{,u}-\frac{2\Lambda}{(D-2)(D-1)}\,r^2 \,.
\label{Hvacuum2}
\ee
Such a metric looks very much like four-dimensional Robinson--Trautman vacuum solutions of type N, see \cite{Stephanibook}. However, for $D>4$ the corresponding Weyl tensor turns out to be either of type D or vanishing (see appendix~A).

As a special subcase of the general $D>4$ Robinson--Trautman class (\ref{geo_metric fin}), (\ref{Hfin}), higher dimensional vacuum spacetimes (with an arbitrary cosmological constant) that admit a hypersurface orthogonal, non-shearing but expanding geodesic null congruence are thus necessarily of the algebraically special type D (cf.~also Appendix~A) or are conformally flat. This is in agreement with Ref.~\cite{Pravdaetal04}, which proved (via a study of the Bianchi identities) that the multiple principal null congruence of $D>4$ type N and type III vacuum spacetimes is geodesic but must have non-zero shear when it has non-zero expansion. For the case of non-twisting null congruences, we have generalized this conclusion to include a non-vanishing cosmological constant, and to prove that any algebraic type different from D and O is forbidden for vacuum spacetimes with an expanding but non-shearing null congruence.

\section{Conclusions}

\label{sec_conclusions}

We have presented the complete family of higher dimensional spacetimes that contain a hypersurface orthogonal, non-shearing and expanding congruence of null geodesics, and that satisfy Einstein's equations with an arbitrary cosmological constant and aligned pure radiation. In particular, we have discussed vacuum solutions. There appear fundamental differences with respect to the standard ${D=4}$ family of Robinson--Trautman solutions. While the latter was originally derived during investigations of exact gravitational waves in general relativity \cite{RobTra60,RobTra62}, it turned out that the vacuum $D>4$ class contains essentially only static black holes. From the viewpoint of the algebraic classification of the Weyl tensor, only the type D (or O in the vacuum case) is permitted, whereas in four dimensions there exist solutions of all special Petrov types \cite{RobTra60,RobTra62,Stephanibook} (with a different approach it was already anticipated in \cite{Pravdaetal04} that type III and type N are forbidden within the class considered here). In addition, even within the type D subclass, there are no higher dimensional counterparts of important $D=4$ Robinson--Trautman spacetimes such as the C-metric, which describes gravitational radiation emitted from black holes in accelerated motion. On the other hand, the $D>4$ Robinson--Trautman class allows for (non-accelerated) black holes with a much richer variety of horizon geometries, namely any compact Einstein space (not necessarily of constant curvature) is permitted.
Consequences of this fact concerning black hole uniqueness and stability where explored, e.g., in \cite{GibIdaShi02prl,GibHar02}.

Along with \cite{Pravdaetal04}, our conclusions indicate that it may be necessary to relax, e.g., the shear-free assumption when looking for expanding radiative solutions in higher dimensions (non-expanding $D>4$ Kundt waves are shear-free and, in this respect, closer to their four-dimensional counterpart, cf.~\cite{ColHerPel06} and references mentioned there). Still within the shearless class considered here, natural generalizations would include additional matter content, such as an electromagnetic field. These developments are left for possible future work.

\section*{Acknowledgments}
%\acknowledgments

The authors are grateful to Jerry Griffiths and Vojt\v ech Pravda
for reading the manuscript. J.P. has been supported by the grant
GA\v{C}R~202/06/0041, M.O. by a post-doctoral fellowship from
Istituto Nazionale di Fisica Nucleare (bando n.10068/03).

\section*{Appendix~A. Weyl tensor and its algebraic type}
\renewcommand{\theequation}{A\arabic{equation}}
\setcounter{equation}{0}

\subsection*{General solution}

For the general Robinson--Trautman line element~(\ref{geo_metric fin}) the only non-vanishing components of the curvature tensor are given by (using eq.~(\ref{constrijs}))
\beqn
 & & R_{ruru}=H_{,rr}\,, \qquad R_{riuj}=r[H_{,r}+(\ln P)_{,u}]h_{ij}\,,  \nonumber \\
 & & R_{ijkl}=r^2{\cal R}_{ijkl}-4r^2\left[H+r(\ln P)_{,u}\right]h_{i[k}h_{l]j} \,,\label{curvature} \\
 & & R_{ruui}=-H_{,ri}+r^{-1}H_{,i}\,, \qquad R_{uijk}=2r^2h_{i[k}(\ln P)_{,j]u}+2rh_{i[k}H_{,j]} \,, \nonumber \\
 & & R_{uiuj}=r\big[[2H+r(\ln P)_{,u}]H_{,r}+H_{,u}-r[(\ln P)_{,u}]^2+r(\ln P)_{,uu}\big]h_{ij} \,, \nonumber\\
  & & \hskip20mm  {}+H_{,ij}-H_{,k}\Gamma^k_{\,ij}\,, \nonumber
\eeqn
where $\Gamma^k_{\,ij}$ and ${\cal R}_{ijkl}$ are, respectively, the Christoffel symbols and the Riemann tensor associated to the spatial metric $h_{ij}$.  When we substitute the explicit form~(\ref{Hfin}) of the function~$H$, we find $R_{ruui}=0=R_{uijk}$. Then, from the standard definition of the Weyl tensor we get
\beqn
 & & C_{ruru}=-\mu(u)\frac{(D-2)(D-3)}{2r^{D-1}} \,, \qquad C_{riuj}=\mu(u)\frac{(D-3)}{2r^{D-3}}h_{ij} \,, \nonumber \\
 & & C_{ijkl}=r^2{\cal R}_{ijkl}-2r^2\left( \frac{{\cal R}(u)}{(D-2)(D-3)}-\frac{\mu(u)}{r^{D-3}}\right)h_{i[k}h_{l]j} \,,\label{Weyl}\\
 & & C_{uiuj}=2H\,C_{riuj}    \nonumber
\eeqn
--- to derive the last component it is necessary to employ the field equations (\ref{constrspec2}), (\ref{hg}) (after differentiating the latter with respect to $u$) and the identities (\ref{christ}), (\ref{diff_scalar}).

Now, using a suitable frame based on the null vectors
\be
 \mbox{\boldmath$k$}=\pa_r , \qquad \mbox{\boldmath$l$}=\pa_u-H\pa_r
 \label{null_vec}
\ee
(such that $\mbox{\boldmath$k$}\cdot\mbox{\boldmath$l$}=-1$), it is straightforward to see that the above coordinate components give raise only to frame components of boost weight 0. The Weyl tensor is thus of type~D according to \cite{Coleyetal04}.
Type O (conformally flat spacetimes) is possible only if $\mu=0$, which implies a vanishing pure radiation field (cf.~eq.~(\ref{constrspec2})). Therefore, as in the case $D=4$ \cite{Stephanibook} {\em there are no conformally flat non-vacuum solutions} in the Robinson--Trautman family.

\subsection*{Vacuum solutions: case $\mu\neq 0$}
For the vacuum line element~(\ref{geo_metric fin}) with the particular function~(\ref{Hvacuum}) the components (\ref{Weyl}) become
\beqn
 & & C_{ruru}=-\mu\frac{(D-2)(D-3)}{2r^{D-1}} \,, \qquad C_{riuj}=\mu\frac{(D-3)}{2r^{D-3}}h_{ij} \,, \nonumber \\
 & & C_{ijkl}=r^2{\cal R}_{ijkl}-2r^2\left( K-\frac{\mu}{r^{D-3}}\right)h_{i[k}h_{l]j} \,,\label{Weylvac1}\\
 & & C_{uiuj}=2H\,C_{riuj}\,.    \nonumber
\eeqn
The Weyl tensor is again of type D.

\subsection*{Vacuum solutions: case $\mu=0$}
For the line element~(\ref{geo_metric fin}),~(\ref{Hvacuum2}) one obtains from  (\ref{Weyl}) that the only non-vanishing components of the Weyl tensor are
\be
 C_{ijkl}=r^2({\cal R}_{ijkl}-2K h_{i[k}h_{l]j})\,.
\label{Weylvac2}
\ee
Such spacetime is again obviously of type D. It degenerates to type O (thus to constant curvature since it is vacuum) when $C_{ijkl}=0$, which is equivalent to having a transverse spatial metric $h_{ij}$ of constant curvature $K$, i.e. a function $P$ of the form~(\ref{flatP}). In particular, this is necessarily the case in $D=5$ (since $h_{ij}$ must always be of constant curvature in that case, see section~\ref{sec_discussion}).

\section*{Appendix~B. Useful identities}
\renewcommand{\theequation}{B\arabic{equation}}
\setcounter{equation}{0}

The relation between the geometries defined by two conformally related metrics $h_{ij}=P^{-2}\gamma_{ij}$ of dimension $D-2$ are  well-known (see, e.g., section~3.7 in \cite{Stephanibook}). The Christoffel symbols $\Gamma^k_{\,ij}$ of $h_{ij}$ are given by
\be
 \Gamma^k_{\,ij}={}^\gamma\Gamma^k_{\,ij}-2\delta^k_{(i}(\ln P)_{,j)}+\gamma_{ij}\gamma^{kl}(\ln P)_{,l} .
 \label{christ}
\ee
The Ricci tensors are related by
\beqn
 {\cal R}_{ij}={}^{\gamma}{\mathcal R}_{ij}+(D-4)\left[(\ln P)_{||ij}+(\ln P)_{,i}(\ln P)_{,j}\right] \nonumber \label{conf_ricci} \\
 {}+\gamma_{ij}\left[{}^{\gamma}\Delta(\ln P)-(D-4)\gamma^{kl}(\ln P)_{,k}(\ln P)_{,l}\right] ,
\eeqn
where $^{\gamma}\Delta$ and the lower double vertical bar denote, respectively, the Laplace operator and the covariant derivative in the geometry of $\gamma_{ij}$. By contraction with $\gamma^{ij}$, one finds
\be
 P^{-2}{\cal R}={}^{\gamma}{\mathcal R}+(D-3)\left[2\,{}^{\gamma}\Delta(\ln P)-(D-4)\gamma^{kl}(\ln P)_{,k}(\ln P)_{,l}\right] .
 \label{conf_scalar}
\ee
For our purposes, it is also useful to observe that differentiating this equation with respect to the ``parameter'' $u$  one obtains (see the main text and recall that ${\gamma_{ij,u}=0}$)
\be
 P^{-2}\left[{\cal R}_{,u}-2{\cal R}(\ln P)_{,u}\right]=(D-3)\left[2\,{}^{\gamma}\Delta(\ln P)-(D-4)\gamma^{kl}(\ln P)_{,k}(\ln P)_{,l}\right]_{,u} .
 \label{diff_scalar}
\ee
Combining eqs.~(\ref{conf_ricci}) and (\ref{conf_scalar}), for the traceless part of the Ricci tensor one gets
\beqn
 {\cal R}_{ij}-\frac{{\cal R}}{D-2}\,h_{ij}={}^{\gamma}{\mathcal R}_{ij}-\frac{{}^{\gamma}{\mathcal R}}{D-2}\,\gamma_{ij}+(D-4)\left[(\ln P)_{||ij}+(\ln P)_{,i}(\ln P)_{,j}\right] \nonumber \label{tracefree} \\
 {}-\frac{D-4}{D-2}\,\gamma_{ij}\left[{}^{\gamma}\Delta(\ln P)+\gamma^{kl}(\ln P)_{,k}(\ln P)_{,l}\right] .
\eeqn

%%\bibliographystyle{JHEP}
%\bibliographystyle{my_cqg}

%\bibliographystyle{elsart-num_mio}
%\bibliography{bibl}

\end{document}